\documentstyle[emulateapj,psfig]{article}

\newcommand\pp     {$\pm$}
\newcommand\pers     {s$^{-1}$}
\newcommand\micros  {$\mu$s}

\def\degr{\hbox{$^\circ$}}

\righthead{Discovery of a $\sim$7 Hz QPO in 4U 1820--30}
\slugcomment{Submitted to ApJ, Nov 1998}

\begin{document}

\title{Discovery of a $\sim$7 Hz Quasi-Periodic Oscillation in
the low-luminosity low-mass X-ray binary 4U 1820--30}

\author{Rudy Wijnands, Michiel van der Klis, \& Erik-Jan Rijkhorst}

\affil{Astronomical Institute ``Anton Pannekoek'' and Center for High
Energy Astrophysics, University of Amsterdam, Kruislaan 403, NL-1098
SJ Amsterdam, The Netherlands; rudy@astro.uva.nl,
michiel@astro.uva.nl, rijkhors@astro.uva.nl}

\begin{abstract}

We have discovered a 7.06\pp0.08 Hz quasi-periodic oscillation (QPO)
in the X-ray flux of the low-luminosity low-mass X-ray binary (LMXB)
and atoll source 4U 1820--30. This QPO was only observable at the
highest observed mass accretion rate, when the source was in the
uppermost part of the banana branch, at a 2--25 keV luminosity of
5.4$\times$10$^{37}$ erg \pers~(for a distance of 6.4 kpc). The QPO
had a FWHM of only 0.5\pp0.2 Hz during small time intervals (32-s of
data), and showed erratic shifts in the centroid frequency between 5.5
and 8 Hz. The rms amplitude over the energy range 2--60 keV was
5.6\%\pp0.2\%. The amplitude increased with photon energy from
3.7\%\pp0.5\% between 2.8 and 5.3 keV to 7.3\%\pp0.6\% between 6.8 and
9.3 keV, above which it remained approximately constant at
$\sim$7\%. The time lag of the QPO between 2.8--6.8 and 6.8--18.2 keV
was consistent with being zero (--1.2\pp3.4 ms).

The properties of the QPO (i.e., its frequency and its presence only
at the highest observed mass accretion rate) are similar to those of
the 5--20 Hz QPO observed in the highest luminosity LMXBs (the Z
sources) when they are accreting near the Eddington mass accretion
limit. If this is indeed the same phenomenon, then models explaining
the 5--20 Hz QPO in the Z sources, which require the near-Eddington
accretion rates, will not hold. Assuming isotropic emission, the 2--25
keV luminosity of 4U 1820--30 at the time of the 7 Hz QPOs is at
maximum only 40\% (for a companion star with cosmic abundances), but
most likely $\sim$20\% (for a helium companion star) of the Eddington
accretion limit.

\end{abstract}


\section{Introduction \label{intro}}

The bright LMXBs can be divided on basis of their correlated X-ray
spectral and timing behavior in the Z sources and the atoll sources,
named after the tracks they produce in the X-ray color-color diagram
(CD; Hasinger \& van der Klis 1989).  The globular cluster source 4U
1820--30 (NGC 6624) was classified as an atoll source. Before the
launch of the {\it Rossi X-ray Timing Explorer} ({\it RXTE}) two types
of QPOs were observed in the Z sources, which were not observed in the
atoll sources (although several indications for QPOs or peaked noise
were found; see, e.g., Stella, White, \& Priedhorsky 1987a; Hasinger
\& van der Klis 1989; Yoshida et al. 1993). On the top branches (the
so-called horizontal branches) of the Z sources a 15--60 Hz QPO was
detected (the horizontal-branch oscillation or HBO) for whose
explanation the magnetospheric beat-frequency model was put forward
(Alpar \& Shaham 1985; Lamb et al. 1985). That the HBO was seen only
in Z sources was thought to be related to a higher magnetic field in
the Z sources compared to that in the atoll sources (Hasinger \& van
der Klis 1989).  In such an interpretation the HBO would be much
weaker in the atoll sources, but they could be observable at some
level.  Another type of QPO was observed on the diagonal branches (the
normal branches) of the Z sources (the normal-branch oscillation or
NBO), when the mass accretion rate is near the Eddington mass
accretion rate (e.g., Hasinger \& van der Klis 1989; Penninx 1989;
Smale 1998). The NBOs were therefore thought to be due to a process
which was activated by such high accretion rates (Fortner et al. 1989;
Alpar et al. 1992), and should be observable in the atoll sources if
they would reach near Eddington accretion rates.

Shortly after the launch of {\it RXTE} in December 1995 a third type
of QPO was discovered in both the Z sources and the atoll sources with
frequencies between 200 and 1200 Hz (the kHz QPOs; see, e.g., van der
Klis et al. 1996; Strohmayer et al. 1996; see Smale, Zhang, \& White
1997 for the discovery of kHz QPOs in 4U 1820--30). In both types of
sources, the kHz QPOs were remarkably similar and were most likely
caused by the same physical mechanism.  In the atoll sources, besided
the kHz QPOs, other QPOs with frequencies between 20 and 70 Hz were
observed (e.g., Strohmayer et al. 1996; Ford et al. 1997; Wijnands et
al. 1998). The properties of these QPOs (their frequencies and their
relation to broad band noise just below the QPO frequency; Wijnands et
al. 1998) suggested a link with the HBOs of the Z sources. So far,
similar QPOs as the Z source NBOs have not been reported in the atoll
sources.

In this Letter, we report the discovery of a QPO near 7 Hz in the
atoll source~4U~1820--30 at its highest observed mass accretion
rate. If this QPO is caused by the same physical mechanism as the NBOs
in the Z sources, then the production mechanism for these QPOs can
also be actived when the luminosity is much less than the critical
Eddington luminosity.

\section{Observations, analysis, and results \label{results}}

\begin{figure*}[t]
\begin{center}
\begin{tabular}{c}
\psfig{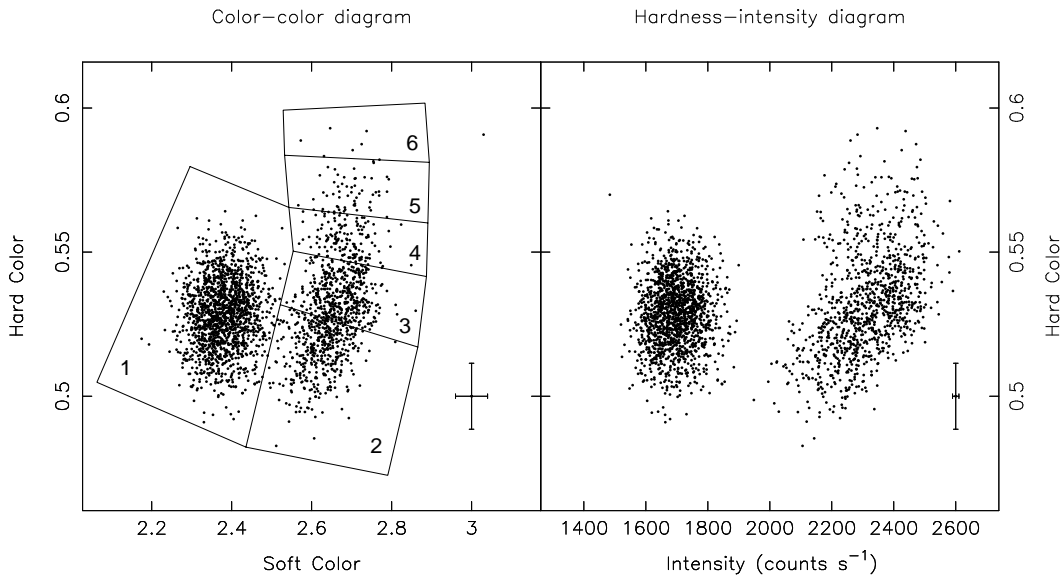}
\end{tabular}
\caption{Color-color diagram ({\it left}) and hardness-intensity
diagram ({\it right}) of 4U 1820--30. The soft color is the
3.5--6.4/2.1--3.5 keV count rate ratio, the hard color is the
9.7--16.0/6.4--9.7 keV ratio, and the intensity is the count rate in
the energy range 2.1--16.0 keV. The colors and the intensity were
calculated for the 3 detectors that were always on. The count rates
were background subtracted, but no dead-time correction was
applied. The dead-time correction is approximately 1\%.  The boxes in
the CD indicate the regions which were used to select the power
spectra.  Typical error bars on the colors and the intensity are
shown.  All data points are 16-s
averages. \label{4u1820-30_nbo_cd_hid}}
\end{center}
\end{figure*}

We used archival data obtained with the {\it RXTE} satellite of 4U
1820--30 on 1996 May 1 09:00--22:51 UT, and 1996 May 3 22:15 -- May 4
18:05 UT. A total of $\sim$40 ks of data were obtained. No X-ray
bursts were observed.  The data were simultaneously
collected with 16 s time resolution in 129 photon energy channels
(covering 2--60 keV), with 2 ms in 16 channels (covering 2--18.2 keV),
and with 16 \micros~in 16 channels (18.2--60 keV). During
$\sim$10\% of the time, detector 4 and/or 5 (out of five) were
off. For the color-color analysis we only used the data of the three
detectors which were always on; for the power spectral analysis we
used all available data.

In order to create the CD and the hardness-intensity diagram (HID), we
used as soft color the 3.5--6.4/2.1--3.5 keV count rate ratio and as
hard color the 9.7--16.0/6.4--9.7 keV ratio. As intensity we used the
count rate in the energy range 2.1--16.0 keV.  The CD and HID of all
data are shown in Figure~\ref{4u1820-30_nbo_cd_hid}.  In the right
hand side of both diagrams what seems to be a banana branch is
observed, with an additional patch of data that might be either the
lower banana branch or part of an island state. The power spectrum
observed during this part (see below) indicates that the source was
not in an island state but in the lower part of the banana branch,
because the strong ($\sim$20\% rms amplitude) band-limited noise,
characteristic of the island state, was not observed. We conclude that
during these observations the source was in the banana branch
continuously.  The gap between the lower part of the banana branch with
the rest is caused by the two day data gap between the 1996 May 1 and
1996 May 3--4 observations. During the May 1 observation the source
was in the lower part of the banana branch and during the May 3--4
observations the source moved up an down the banana branch.

\begin{figure*}[t]
\begin{center}
\begin{tabular}{c}
\psfig{figure=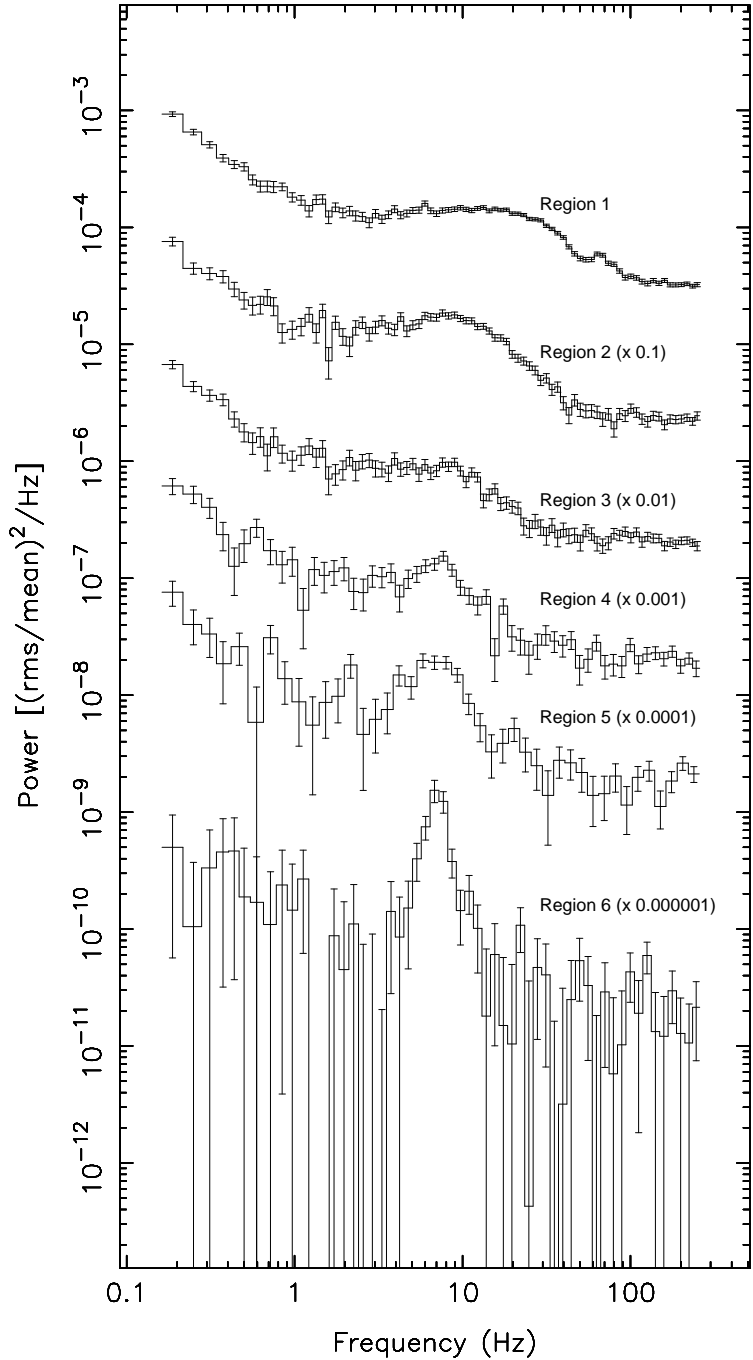,width=10cm}
\end{tabular}
\caption{Power spectra of 4U 1820--30 corresponding to the different
regions in the CD indicated in Figure~\ref{4u1820-30_nbo_cd_hid}. The
power spectra appear flat at high frequencies, as for display
purposes, only 96\% of the dead-time modified Poisson level was
subtracted. \label{4u1820-30_nbo_cd_powerspectra}}
\end{center}
\end{figure*}

In order to study the timing behavior as a function of the position of
4U 1820--30 on the banana branch, we made power spectra of 16-s data
segments of the 2 ms and 16 \micros~data combined. This resulted in
$\sim$2500 power spectra between 1/16 and 256 Hz.  The rms amplitude
values quoted are those obtained for the energy range 2--60 keV,
unless otherwise stated.  We selected the power spectra on basis of
the position of the source in the CD by dividing the banana branch in
the CD into six regions (Figure~\ref{4u1820-30_nbo_cd_hid} {\it
left}). Each region was analyzed separately.  The power spectra
corresponding to each region were averaged and the dead-time modified
Poisson level (Zhang 1995; Zhang et al. 1995) was subtracted.  The
average power spectra of regions 2 and 3 were fitted with a fit
function consisting of a power law (representing the very-low
frequency noise [VLFN]) and an exponentially cutoff power law
(representing the high-frequency noise [HFN]). For region 1 a
Lorentzian was included in the fit function as well, to fit a QPO near
70 Hz. For the power spectra corresponding to the other three regions,
we only used a power law for the VLFN and a Lorentzian for the 7 Hz
QPO (region 6) or the broad bump near 7 Hz (regions 4 and 5).

The power spectra corresponding to each region are shown in
Figure~\ref{4u1820-30_nbo_cd_powerspectra}.  In the lower banana
branch (region 1) the atoll source VLFN and HFN components (see
Hasinger \& van der Klis 1989) were observed.  A QPO at 71\pp1 Hz was
also observed, with an rms amplitude of 2.2\%\pp0.2\% and a FWHM of
21\pp4 Hz.  There are also indications for a broad noise components at
$\sim$100 Hz, similar to what has found in other atoll sources (Ford,
van der Klis, \& Kaaret 1998; Ford \& van der Klis 1998; Wijnands \&
van der Klis 1998a), however, the limited time resolution below 18.2
keV did not allow to investigate this in more detail. The VLFN was
detected throughout the entire banana branch except in region 6.  Its
rms amplitude (integrated over 0.01--1 Hz) varied erratically between
2.7\% and 7.9\%, with a 95\% confidence upper limit in region 6 of
3.4\%. The power law index varied erratically between 1.0 and 1.7.
The HFN was observed in regions 1--3. Its rms amplitude (integrated
over 1--100 Hz) and cutoff frequency decreased from 6.3\%\pp0.1\% and
13.7\pp0.7 Hz in region 1, to 3.1\%\pp0.1\% and 5.5\pp0.6 Hz in region
3. The power law index varied erratically between $-$0.9 and
$-$1.4. In regions 4 and 5, broad peaked noise is present near 7
Hz. It is unclear whether the HFN evolved into this broad peaked noise
component or was replaced by it. In region 6, a very significant
(9$\sigma$) QPO is present near 7 Hz, which might be related to the
peaked noise in region 4 and 5. We fitted this peaked noise as well as
the QPO with a Lorentzian and found frequencies of 7.4\pp0.3 Hz in
region 4 and 7.1\pp0.1 Hz in region 5, consistent with the QPO
frequency of 7.06\pp0.08 Hz in region 6. While the FWHM of the
Lorentzian decreased from 6.4\pp1.8 Hz in region 4 to 1.3\pp0.3 Hz in
region 6, the rms amplitude increased from 3.4\%\pp0.4\% to
6.2\%\pp0.3\%.

\begin{figure*}[t]
\begin{center}
\begin{tabular}{c}
\psfig{figure=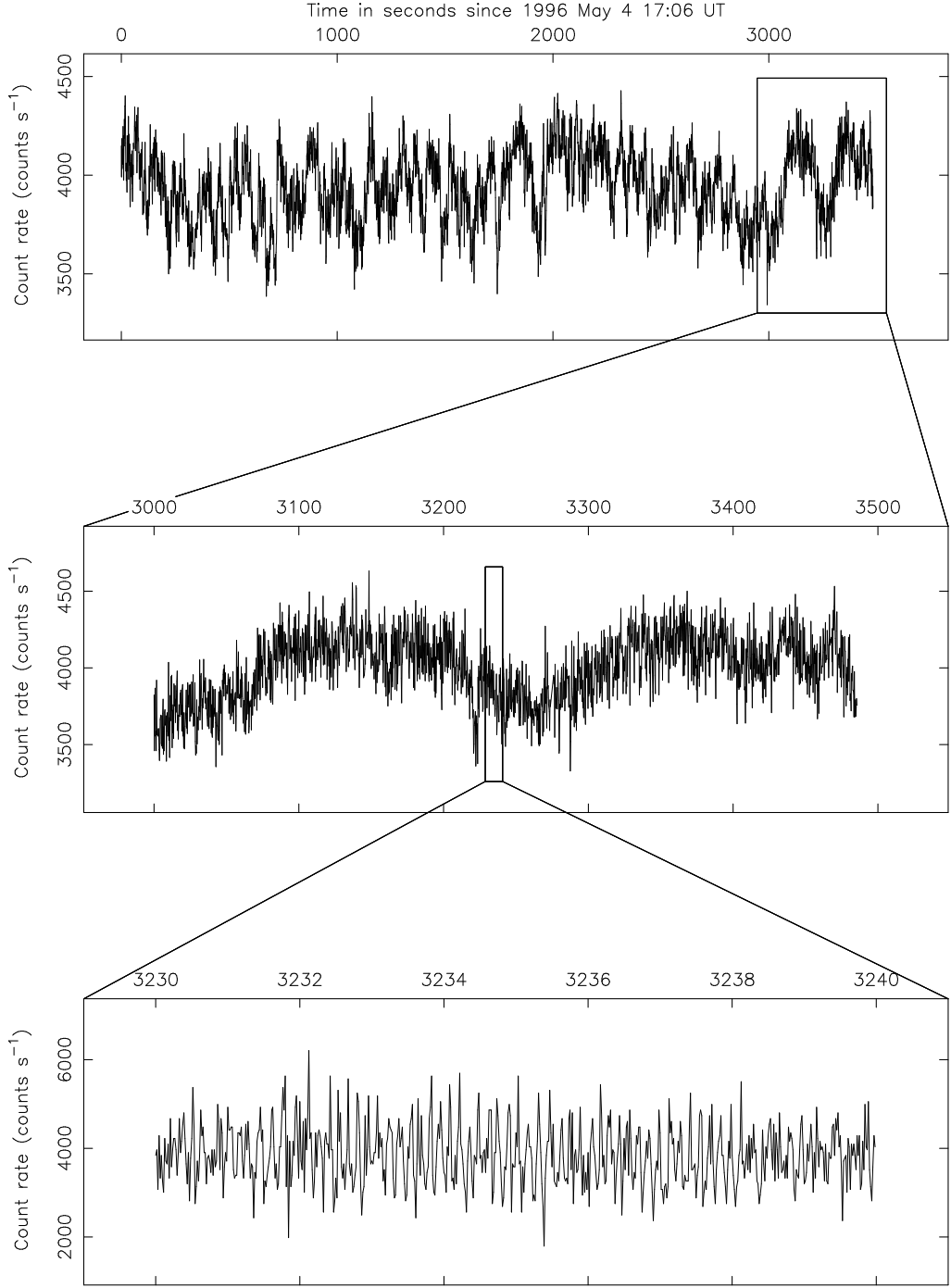,width=14cm,angle=180}
\end{tabular}
\caption{The 2--18.2 keV light curves of 4U 1820--30 on 1996 May 4. In
the top panel each point corresponds to 1 s, in the middle panel to
0.25 s, and in the bottom panel to 1/64 s. Typical errors are 70
counts~\pers~in the top panel, 125 counts~\pers~in the middle panel,
and 500 counts~\pers~in the bottom panel. The QPO is only visible
between $\sim$3150 and 3350 s since the start of the
observation.\label{4u1820-30_nbo_lightcurve}}
\end{center}
\end{figure*}

We investigated the presence of the 7 Hz QPO in more detail and we
found that it was only present during approximately 160 s of data
during the 1996 May 4 observation.  The QPO could then be directly
observed in the light curve (Figure~\ref{4u1820-30_nbo_lightcurve}
{\it bottom panel}). When measured in 32-s time intervals the QPO
frequency shifted erratically between 5.5 and 8 Hz and had at certain
times a FWHM of only 0.5\pp0.2 Hz. The frequency drifts of the QPO
caused the FWHM to be $\sim$2~Hz in all QPO data combined. We used
these 160 s of data showing the QPO to study the energy dependence of
its amplitude. We fixed the frequency (7.2\pp0.1 Hz) and FWHM
(2.2\pp0.4 Hz) to the values obtained by using the total energy range
(rms amplitude of 5.6\%\pp0.2\%; 11.8$\sigma$). The rms amplitude in
the energy bands 2.8--5.3, 5.3--6.8, 6.8--9.3, 9.3--10.8, and
10.8--18.2 keV, were 3.7\%\pp0.5\%, 5.0\%\pp0.8\%, 7.3\%\pp0.6\%,
7.1\%\pp1.3\%, and 6.9\%\pp0.9\%, respectively.  The 95\% confidence
upper limits on the presence of the QPO outside the 160 s interval
were between 1\% and 1.5\% (total energy range).  We performed time
lag measurements between the energy bands 2.8--6.8 and 6.8--18.2
keV. The time lag of --1.2\pp3.4 ms is consistent with being zero.

We determined the X-ray flux during the time of the QPO by fitting the
X-ray spectrum with an absorbed black body plus a power law.  We
obtained a 2--25 keV X-ray flux of 1.1$\times$10$^{-8}$ erg
\pers~cm$^{-2}$. This corresponds to a 2--25 keV X-ray luminosity of
5.4$\times$10$^{37}$ erg \pers, assuming a distance of 6.4 kpc (Vacca,
Lewin, \& van Paradijs 1986). The power law index was $\sim$2.8
indicating that not much flux was present above 25 keV.

\section{Discussion \label{discussion}}

We have discovered a very significant $\sim$7 Hz QPO in the X-ray flux
of the globular cluster LMXB and atoll source 4U 1820--30. This QPO
was only observed during a small part of the data, when the source was
in the uppermost part of the banana branch in the CD. It is thought
that the mass accretion rate increases monotonically along the banana
branch from the lower to the upper banana (Hasinger \& van der Klis
1989). If true, then the 7 Hz QPO was only detected during the highest
observed mass accretion rate in this source. Note that, although the
QPO was only detectable when the source was at the uppermost part of
the banana branch, the count rate then was not the highest observed
(Fig.~\ref{4u1820-30_nbo_lightcurve}). It seems that in this source,
similar to~what has been observed in the atoll sources 4U 1636--53
(van der Klis 1990; Prins \& van der Klis 1997) and 4U 1608--52
(M\'endez et al. 1998), the count rate does not have a one-to-one
relation with the mass accretion rate.

Owing to the small amount of data showing the QPO it is difficult to
compare this QPO with QPOs observed in other X-ray binaries. Similar
frequency QPOs have been observed in a wide range of sources. For
example, in black hole candidates (BHCs) QPOs with frequencies around
7 Hz were observed in several systems (see van der Klis 1995 for a
review). However, recently it was shown (Wijnands \& van der Klis
1998b) that at least some~of~these BHC QPOs are most likely related to
the QPOs seen in atoll sources between 20 and 70 Hz. When 4U 1820--30
is on the lower part of the banana branch, thus at lower mass
accretion rates, we detect a significant $\sim$70 Hz QPO
(Fig.~\ref{4u1820-30_nbo_cd_powerspectra}), similar to what is
observed in other atoll sources. This indicates that the 7 Hz QPO is
most likely not related to the 7 Hz BHC QPOs, unless the 70 Hz QPO
decreased in frequency from 70 to 7 Hz from the lower banana branch to
the upper part, opposite to what would be expected (e.g., Ford \& van
der Klis 1998).

In the X-ray dipper 4U 1323--62 a strong ($\sim$10\% rms amplitude)
QPO was recently discovered near 1 Hz (Jonker et al. 1998). The
production of this QPO is thought to be related to the high
($>$70\degr) inclination of this system. Although the modulation of
the X-ray flux at the 11-min orbital period (Stella, Priedhorsky, \&
White 1987b) suggested a high inclination for 4U 1820--30, the strong
modulation observed in the ultraviolet flux of this source indicates
that the inclination most likely is between 35\degr~and
50\degr~(Anderson et al. 1997), making a common production mechanism
of the two types of QPOs unlikely. This conclusion is further
strengthened by the differences between the QPOs.  The strength of the
1 Hz QPO in 4U 1323--62 only marginally increased with photon energy
and it is observed almost continuously (even in the dips and
the X-ray bursts), both contrary to the case of our 7 Hz QPO. 

The most likely identification of the 7 Hz QPO in 4U 1820--30 is with
the 7 Hz QPOs observed in the Z sources, when they are accreting near
the Eddington mass accretion rate (see van der Klis 1995).  This
identification cannot be tested by comparing the time lags of the
different QPOs. The non-detection of time lags in 4U 1820--30 is
consistent with the non-detection in Scorpius X-1 (Dieters et
al. 1999), but inconsistent with the large ($\sim$80 ms) time lags
reported in Cygnus X-2 (Mitsuda \& Dotani 1989).  However, besides the
similar frequencies, also the power spectral evolution with increasing
mass accretion rate is very similar between the Z sources and 4U
1820--30. In the Z sources at low observed accretion rates, relatively
strong (10--20\%) band limited noise is present, which decreases in
amplitude when the accretion rate is increasing. At a certain
accretion rate, this noise is hardly detectable, but a new component
arises near 7 Hz, which evolves in a clear QPO. The power spectral
evolution of 4U 1820--30 is remarkably similar
(Fig.~\ref{4u1820-30_nbo_cd_powerspectra}).  Seen in this light, it is
not surprising that the 7 Hz QPO in 4U 1820--30 is only observed in
the highest observed mass accretion rate. The NBOs in the Z sources
are observed when these sources accretion near the Eddington accretion
rate (Hasinger \& van der Klis 1989; see, e.g., Penninx 1989 and Smale
1998 for the value of the accretion rate when the Z sources are on
their normal branches). However, the X-ray luminosity in 4U 1820--30
is, assuming isotropic emission, at maximum $\sim$40\% of the
Eddington critical luminosity if the companion star has cosmic
abundance.  As the companion star is most likely a helium white dwarf
(Stella et al. 1987a), the X-ray luminosity might be only $\sim$20\%
of the (helium) Eddington critical luminosity. Thus, barring a large
anisotropy in the X-ray emission, the accretion rate in 4U 1820--30 is
well below the Eddington critical accretion rate and below the
accretion rate of the Z sources when they exhibited their 7 Hz QPOs.
This would mean that, if the 7 Hz QPO in 4U 1820--30 is caused by the
same physical mechanism as similar frequency QPOs in the Z sources,
then the formation mechanism behind these QPOs is already activated in
4U 1820--30 well below the critical Eddington luminosity.

Note added in manuscript: After we submitted this Letter, we analyzed
the PCA data of 4U 1820--30 used by Smale et al. (1997) and Zhang et
al. (1998). During some of their observations the 7 Hz QPO was
detected, again only at times when the source was in the uppermost
part of the banana branch.  When the source was at slightly lower
inferred mass accretion rate, broad peaked noise was present near 10
Hz, which might be related to the QPO. These observations will be
discussed in a forthcoming paper.

\acknowledgments

This work was supported in part by the Netherlands Foundation for
Research in Astronomy (ASTRON) grant 781-76-017.  This research has
made use of data obtained through the High Energy Astrophysics Science
Archive Research Center Online Service, provided by the NASA/Goddard
Space Flight Center. We thank Jeroen Homan for stimulating discussions
and comments on an earlier version of this document.

\end{document}